\documentclass[12pt]{article}

\begin{document}
\renewcommand{\thefootnote}{\fnsymbol{footnote}}

\begin{titlepage}
%---------------- preprint number & date ---------------
%\hfill\parbox{4cm}{hep-th/0206090}
\hfill{hep-th/0206090}
%------------------------ title ------------------------
\vspace{15mm}
\baselineskip 8mm

\begin{center}
{\LARGE \bf Branes from Matrix Theory \\
in PP-Wave Background }
\end{center}
\baselineskip 6mm
%---------------- authors and addresses ----------------
\vspace{10mm}
\begin{center}
Seungjoon Hyun$^a$\footnote{\tt hyun@phya.yonsei.ac.kr} and
Hyeonjoon Shin$^b$\footnote{\tt hshin@newton.skku.ac.kr} \\[5mm]
{\it
$^a$Institute of Physics and Applied Physics, Yonsei University,
Seoul 120-749, Korea \\
$^b$ BK 21 Physics Research Division and Institute of Basic Science \\
Sungkyunkwan University, Suwon 440-746, Korea}
\end{center}

\thispagestyle{empty}

%----------------------- abstract ----------------------

\vfill

\begin{center}
{\bf Abstract}
\end{center}
\noindent Based on the recently proposed action for Matrix theory
describing the DLCQ M theory in the maximally supersymmetric
pp-wave background, we obtain the supersymmetry algebra of
supercharge density.  Using supersymmetry transformation rules for
fermions, we identify BPS states with the central charges in the
supersymmetry algebra, which can be activated only in the large
$N$ limit. They preserve some fraction of supersymmetries and
correspond to rotating transverse membranes and longitudinal five
branes.

\vspace{20mm}
\end{titlepage}
%-------------------------------------------------------

\baselineskip 6.5mm
\renewcommand{\thefootnote}{\arabic{footnote}}
\setcounter{footnote}{0}

%-------------------- Body of paper --------------------

\section{Introduction}

The matrix theory has been a candidate for the microscopic
description of the eleven dimensional M theory in the infinite
momentum \cite{ban043} or light cone frame
\cite{sus080,sei009,sen220} (or under the discrete light cone
quantization (DLCQ)).  Though we are still far from the complete
understanding of M theory, the matrix theory has been successful
in uncovering some aspects of M theory. For a review, see
Ref.~\cite{tay126}.  In view of its successes, quite a number of
supersymmetries has played a crucial role in exploring the matrix
theory.\footnote{As a typical example, see \cite{hyu022}.}

In its original form, the matrix theory is for the DLCQ M theory
on the flat eleven dimensional Minkowskian space-time of maximal
32 supersymmetries as the background geometry.  Recently, it has
been proposed in Ref.~\cite{ber021} that the original matrix
theory is a special case of a newly constructed matrix theory
which is interpreted as the description of DLCQ M theory on the
eleven dimensional pp-wave background.  The pp-wave geometry
\cite{kow194,guv061,fig308} has 32 supersymmetries and the
limiting case of the eleven dimensional $AdS$ type geometries
under the Penrose limit \cite{pen271} as shown in \cite{bla081}.
In our notation, it is given by
\begin{eqnarray}
& & ds^2 = - 2 dx^+ dx^-
    - \left( \sum^3_{i=1} \frac{\mu^2}{9} (x^i)^2
            +\sum^9_{i'=4} \frac{\mu^2}{36} (x^{i'})^2
      \right) (dx^+)^2
    + \sum^9_{I=1} (dx^I)^2~,
                                      \nonumber \\
& & F_{+123} = \mu~,
\label{pp-wave}
\end{eqnarray}
where $\mu$ is a parameter of the geometry and characterizes the
matrix theory of Ref.~\cite{ber021}.  The matrix theory based on
this geometry has different structure compared to the original
matrix theory and hence opens the possibility for us to get some
more insights for the structure of the M theory.  In particular,
the fact that the geometry (\ref{pp-wave}) is maximally
supersymmetric means that the full power of supersymmetry can
still be utilized as in the case of the flat space-time.

The action for the matrix theory proposed in Ref.~\cite{ber021}
has been constructed starting from the action for the eleven
dimensional superparticles on the above geometry and requiring
supersymmetry.  In a subsequent work, \cite{das185}, the same
action has been obtained through the light cone gauge formulation
of the eleven dimensional super membrane in the pp-wave geometry
by using the action given in \cite{dew209,cla045}. The resulting
theory depends on the parameter $\mu$ which makes the theory to be
massive and has cubic interaction term called the Myers term
\cite{mye053}. Interestingly, the presence of the cubic
interaction allows one to have 1/2 BPS spherical membrane even for
{\em finite} $N$ \cite{ber021,das185}. Here $N$ measures the light
cone momentum of DLCQ M theory and the dimension of the matrix
theory quantities.

For finite $N$, the spherical membrane and its dialects may be the
only extended objects present in the matrix theory.  However,
there may be other types of supersymmetric objects which are
infinitely extended and thus activated only when $N$ goes to
infinity.  In the flat case, one can read off the presence of the
extended objects by looking at the supecharge density algebra
\cite{ban157}.  The central charges are identified as the charges
of the corresponding extended objects.  In this paper, following
the recipes given in \cite{ban157}, we will obtain the supercharge
density algebra of the matrix theory of \cite{ber021,das185} and
investigate what types of extended objects appear from the theory.
In fact, the supersymmetry algebra has already been given in
Ref.~\cite{das185}, which is however valid only for finite $N$. We
also find various rotating BPS brane solutions, which exists only
in the large $N$ limit. They include transverse membranes
stretched in the $x^{i'}$-directions, longitudinal five branes
spanning in the $x^{i'}$-directions and longitudinal five branes
filling 1, 2, 3-directions and spanning one direction in $x^{i'}$.

The organization of this paper is as follows.  In section
\ref{alg}, we will compute the supercharge density algebra. In
section \ref{brane}, we will obtain various rotating transverse
membranes and longitudinal five branes. In section \ref{conc} we
draw some conclusions.

\section{Supercharge Density Algebra}
\label{alg}

The Lagrangian \cite{ber021,das185} for the matrix theory in the
pp-wave background Eq.~(\ref{pp-wave}) is given
by
\begin{equation}
L = {\rm Tr} {\cal L}~.
\label{lag}
\end{equation}
${\cal L}$ is the Lagrangian density in the sense that it is not
traced, and its expression is
\begin{eqnarray}
{\cal L}
 &=& \frac{1}{2R} D_0 X^I D_0 X^I
 -\frac{1}{2R} \left( \frac{\mu}{3} \right)^2 (X^i)^2
 -\frac{1}{2R} \left( \frac{\mu}{6} \right)^2 (X^{i'})^2
 + \frac{i}{2} \theta^\alpha D_0 \theta^\alpha
 - \frac{i\mu}{8} \theta^\alpha
        \Pi_{\alpha \beta} \theta^\beta
                    \nonumber \\
 & & + \frac{R}{4} [ X^I, X^J ]^2
 + \frac{R}{2} \theta^\alpha
      \gamma^I_{\alpha\beta}
      [ X^I, \theta^\beta ]
 - \frac{i\mu}{3} \epsilon_{ijk} X^i X^j X^k~,
\end{eqnarray}
where $D_0 = \partial_0 - i [A_0, ]$ and we have defined
\begin{equation}
\Pi = \gamma^{123}~.
\end{equation}
$R$ is the radius of the longitudinal $x^-$ direction. $X^I$,
$\theta^\alpha$ ($\alpha=1,...,16$), and $A_0$ are hermitian $N \times
N$ matrices which have upper and lower $SU(N)$ indices ($a,b,... =
1,...,N$).  The multiplication between matrices is then given by
\begin{equation}
(AB)^b_a = A^c_a B^b_c~.
\end{equation}

The matrix theory is supersymmetric and, by definition, describes the
eleven dimensional M theory in the infinite momentum or light cone
frame.  In the light cone frame, it is natural to split the eleven
dimensional 32 supersymmetries into 16 kinematical and 16 dynamical
supersymmetries.  If we take $A_0=0$ gauge, the corresponding
transformation rules ($\tilde{\delta}$) for the former one are
\begin{eqnarray}
& &\tilde{\delta} X^I = 0~,
                    \nonumber \\
& &\tilde{\delta} \theta= \tilde{\eta}~, \label{kin}
\end{eqnarray}
where, unlike the flat case, the transformation parameter depends
on time and, with constant $\tilde{\epsilon}$, is as follows:
\begin{equation}
\tilde{\eta} = e^{\frac{\mu}{4} \Pi t} \tilde{\epsilon}~.
\end{equation}
As for the dynamical supersymmetries, the transformation rules
($\delta$) are
\begin{eqnarray}
& & \delta X^I = i \theta \gamma^I \eta~,
                    \nonumber \\
& & \delta \theta  =
  \left(
     P^I \gamma^I
  + \frac{i}{2} [X^I, X^J] \gamma^{IJ}
  + \frac{\mu}{3R} X^i \Pi \gamma^i
  - \frac{\mu}{6R} X^{i'} \gamma^{i'}\Pi
  \right) \eta~,
  \label{dyn}
\end{eqnarray}
where $P^I$ is the canonical conjugate momentum of $X^I$, and
$\eta$ has different time dependence from that of $\tilde{\eta}$
as
\begin{equation}
\eta = e^{- \frac{\mu}{12} \Pi t } \epsilon~,
\end{equation}
with constant $\epsilon$.

The Lagrangian (\ref{lag}) has fermionic sector which is first order
in time derivative.  Thus the conjugate momentum of fermion gives
constraint which is in second class.  Before we compute the
supersymmetry algebra, we should first take it into account properly.
First of all, we obtain the Hamiltonian corresponding to the
Lagrangian, which in the $A_0=0$ gauge is
\begin{equation}
H = R {\rm Tr} {\cal H}~.
\end{equation}
Here the Hamiltonian density ${\cal H}$ is given by
\begin{eqnarray}
{\cal H}
 &=& \frac{1}{2} (P^I)^2
   +\frac{1}{2} \left( \frac{\mu}{3R} \right)^2 (X^i)^2
   +\frac{1}{2} \left( \frac{\mu}{6R} \right)^2 (X^{i'})^2
 - \frac{1}{4} [ X^I, X^J ]^2
                                        \nonumber \\
 & & + i \frac{\mu}{3R} \epsilon_{ijk} X^i X^j X^k
     + i\frac{\mu}{8R} \theta^\alpha \Pi_{\alpha\beta}
                       \theta^\beta
     + \frac{1}{4} \gamma^I_{\alpha\beta} [ \theta^\alpha, [
     \theta^\beta, X^I ]]~.
\end{eqnarray}
If we investigate the constraint structure with this Hamiltonian, we
see that no more constraint is generated. The Dirac procedure for the
constrained system then leads us to the following Dirac brackets for
the elementary matrix quantities:
\begin{eqnarray}
& & \{ X^{I b}_a, P^{J d}_c \}_{DB}
    = \delta^{IJ} \delta^d_a \delta^b_c~,
                    \nonumber \\
& & \{ \theta^{\alpha b}_a, \theta^{\beta d}_c \}_{DB}
    = -i \delta^{\alpha \beta} \delta^d_a \delta^b_c~.
    \label{db}
\end{eqnarray}
From now on, we will keep the subscript $DB$ to distinguish the Dirac
bracket from the commutator and the anti-commutator between matrices.

We now turn to the calculation of the algebra between supercharge
densities.  We denote $Q_\alpha$ ($\tilde{Q}_\alpha$) the 16 dynamical
(kinematical) supercharges that generate the dynamical (kinematical)
supersymmetry transformation $\delta$ ($\tilde{\delta}$). Then the
supercharges for the matrix theory in the pp-wave background are
\begin{eqnarray}
 & & Q_\alpha = {\rm Tr} q_\alpha~, \nonumber \\
 & & \tilde{Q}_\alpha = {\rm Tr} \tilde{q}_\alpha~,
\end{eqnarray}
where the supercharge densities $q_\alpha$ and $\tilde{q}_\alpha$
are $N\times N$ matrices and given by
\begin{eqnarray}
& & q_{\alpha a}^b = \sqrt{\frac{R}{2}} \bigg\{ P^I
\gamma^I_{\alpha \alpha'} - \frac{i}{2} [ X^I, X^J]
\gamma^{IJ}_{\alpha \alpha'} - \frac{\mu}{3R} X^i (\Pi
\gamma^i)_{\alpha \alpha'} + \frac{\mu}{6R} X^{i'} (\Pi
\gamma^{i'})_{\alpha \alpha'}, \theta^{\alpha'} \bigg\}^b_a~,
                                        \nonumber \\
& & \tilde{q}_{\alpha a}^b = \sqrt{\frac{2}{R}} \delta_{\alpha
\alpha'} \theta^{\alpha' b}_a~.
\end{eqnarray}
The purpose of computing the Dirac brackets between supercharge
densities is to see the central charges of extended objects which
disappear for finite $N$ if we trace the matrix indices.  However, the
actual calculation is fairly complicated and hence some simplification
is required.  As pointed out in Ref.~\cite{ban157}, if we drop pieces
of the resulting answer which are antisymmetric in the spinor indices,
we can freely trace on one of the terms in the Dirac bracket.  This
simplifies and reduces the calculation and furthermore does not touch
the structure of the central charge densities. What we are doing is
thus the computation of Dirac brackets between supercharge density and
supercharge.

In the process of calculation, we need the following $SO(9)$ Fierz
identities.
\begin{eqnarray}
& &\theta_\alpha \theta_\beta = \frac{1}{32} (\gamma^I \gamma^J
)_{\alpha \beta} \theta \gamma^I \gamma^J \theta + \frac{1}{96}
(\gamma^I \gamma^J \gamma^J)_{\alpha \beta} \theta \gamma^I
\gamma^J \gamma^K \theta~,
                                \nonumber \\
& & \gamma^I_{\alpha \alpha'} \gamma^{IJ}_{\beta \beta'} +
\gamma^{IJ}_{\alpha \alpha'} \gamma^I_{\beta \beta'}+ (\alpha
\leftrightarrow \beta) = 2 \gamma^J_{\alpha' \beta'}
\delta_{\alpha \beta} - 2 \gamma^J_{\alpha \beta} \delta_{\alpha'
\beta'}~.
\label{fierz}
\end{eqnarray}
Using these identities and the Dirac brackets, Eq.~(\ref{db}), we
obtain after slightly tedious manipulation
\begin{eqnarray}
\{ \tilde{q}_{\alpha a}^b, \tilde{Q}_\beta \}_{DB}
 &=& - i \frac{2}{R} \delta_{\alpha \beta} \delta^b_a~,
                                    \nonumber \\
\{ q_{\alpha a}^b, \tilde{Q}_\beta \}_{DB}
 &=& -2 i \left( P^I \gamma^I_{\alpha \beta}
     - \gamma^{IJ}_{\alpha \beta} z^{IJ}
    - \frac{\mu}{3R} X^i (\Pi \gamma^i)_{\alpha\beta}
    + \frac{\mu}{6R} X^{i'} (\Pi \gamma^{i'})_{\alpha\beta}
    \right)^b_a~,
                                    \nonumber \\
\{ q_{a (\alpha}^b, Q_{\beta)} \}_{DB}
 &=& -4i R {\cal H}_a^b \delta_{\alpha\beta}
   +i \frac{2\mu}{3} (\Pi \gamma^{ij} )_{\alpha\beta} J^{ijb}_a
   -i \frac{\mu}{3} (\Pi \gamma^{i'j'})_{\alpha\beta}J^{i'j'b}_a
                                    \nonumber \\
 & &
   -2i \gamma^I_{\alpha\beta} z_a^{Ib}
   -2i \gamma^{IJKL}_{\alpha\beta} z_a^{IJKLb}
                                    \nonumber \\
 & & - \frac{\mu}{3} (\Pi \gamma^{j'})_{\alpha\beta}
     [ 2(X^i)^2 - (X^{i'})^2, X^{j'} ]^b_a
                                    \nonumber \\
 & &
     - \frac{\mu}{6} \epsilon_{ijk} \gamma^{iji'j'}_{\alpha\beta}
     [ X^{i'}, \{ X^k, X^{j'} \} ]^b_a~.
\label{susy-alg}
\end{eqnarray}
where we have symmetrized over $\alpha$ and $\beta$ in the last
Dirac bracket.  Some definitions and remarks are as follows:
$J^{ij}$ and $J^{i'j'}$ are $SO(3)$ and $SO(6)$ rotation
generators respectively, which are given by
\begin{eqnarray}
J^{ij} &=&  X^i P^j - P^i X^j
          - \frac{i}{4} \theta \gamma^{ij} \theta~,
                                    \nonumber \\
J^{i'j'} &=& X^{i'} P^{j'} - P^{i'} X^{j'}
           - \frac{i}{4} \theta \gamma^{i'j'} \theta~.
\label{ang}\end{eqnarray} The $z$'s represent the central charge
densities for the matrix theory in the flat case \cite{ban157} and
are given by
\begin{eqnarray}
z^I &=& i R \{ P^J, [ X^J, X^I ]]
 - \frac{R}{2} [ \theta^{\alpha'}, [ \theta^{\alpha'}, X^I ]]
                                    \nonumber \\
z^{IJ} &=& \frac{i}{2} [ X^I, X^J ]
                                    \nonumber \\
z^{IJKL} &=& R X^{[I} X^J X^K X^{L]}
\end{eqnarray}
These charges are interpreted as those of the wrapped membrane,
membrane, and wrapped (or longitudinal) fivebrane, respectively.  The
wrapped objects extend along the light cone $x^-$ direction and hence
we have $R$ dependence in the corresponding expressions for them.  For
finite $N$, they do not activate when we take matrix trace.  The
vanishing of ${\rm Tr} z^{IJ}$ and ${\rm Tr} z^{IJKL}$ for finite $N$
is obvious.  For the central charge $z^I$, as noted in
Ref.~\cite{ban157}, we should use the Gauss law constraint $\Phi$
which is obtained as
\begin{equation}
\Phi_a^b = \frac{\partial L }{\partial A_{0b}^a} = i[ P^I, X^I
]_a^b - (\theta^\alpha \theta^\alpha)_a^b
\end{equation}
In terms of this constraint, $z^I$ can be rewritten as
\begin{equation}
z^I  = R \{ \Phi, X^I \} + i R [ X^J, \{ P^J, X^I \} ]
 + \frac{R}{2} \{ \theta^\alpha, \{ \theta^\alpha, X^I \}\}
\end{equation}
In the $A_0=0$ gauge, $\Phi=0$ and ${\rm Tr} z^I$ vanishes for
finite $N$.  As a final remark, we would like to note that, for
finite $N$, the algebra (\ref{susy-alg}) reduces to that obtained
in \cite{das185} or the super pp-wave algebra \cite{hat002} in the
context of the eleven dimensional supergravity.

\section{BPS Branes in the Matrix Model on pp-wave Background}
\label{brane}

One can read off the possible BPS solutions from the supersymmetry
transformation rules (\ref{kin}) and (\ref{dyn}). The only
possible static solution one can get is the fuzzy sphere in the
$x^i$-directions\cite{ber021}
\begin{equation}
[ X^i, X^j ] = i {\mu \over 3R}\epsilon_{ijk} X^k
\end{equation}
The kinematical supersymmetry generators are completely broken
while all the dynamical supersymmetry generators are preserved.
Thus it is half-BPS state and describe spherical membrane. This
configuration is given in \cite{ber021} and extensively discussed
in \cite{das185}.

Our focus is to find all the BPS solutions including the
non-static branes.\footnote{Some BPS solutions related to this has
been given in \cite{bak033}.} There exist non-static flat membrane
solutions spanning in $X^{ i'}$ directions.  They are given by the
configurations
\begin{eqnarray}
X_{4}=r_1 \cos(\mu t/6)~, \ \ \ X_{7}=r_1 \sin(\mu t/6)~,\cr
X_{5}=r_2 \cos(\mu t/6)~,\ \ \ X_{8}=r_2 \sin(\mu t/6)~,
\label{conf1-1}
\end{eqnarray}
where
\begin{equation}
[r_1, r_2]= i {\cal F}_{12}~,\label{conf1-2}
\end{equation}
where $I$ is the unit matrix in the $SU(N)$ space and ${\cal
F}_{ab}$ is antisymmetric. The solutions exist only in the large
$N$ limit as discussed in the previous section.

This configuration has $\frac{1}{8}$ supersymmetry and corresponds
to the rotating transverse  membrane expanded along $X^{i'}$
directions with angular frequency $\omega=\frac{\mu}{6}$. The four
dynamical supercharges which preserve the above configuration are
those satisfying
\begin{eqnarray}
\Pi \epsilon=  \gamma_{47}\epsilon =  \gamma_{58}\epsilon = \pm
i\epsilon~,~.
\end{eqnarray}
in which the transformation rule (\ref{dyn}) becomes
\begin{eqnarray}
\delta\theta &=& \frac{i}{2} [ r_1, r_2]
 e^{\pm{i\mu \over 4} t}\gamma_{45}\epsilon~.
  \label{dyn-c}
\end{eqnarray}
These are cancelled with the kinematical supersymmetry
transformations generated by ${\tilde \epsilon}$ satisfying
\begin{eqnarray}
\Pi {\tilde \epsilon}=  \pm i{\tilde \epsilon}~,~.
\end{eqnarray}
and thus given by
\begin{eqnarray}
{\tilde \delta}\theta =
 e^{\pm{i\mu \over 4} t}{\tilde \epsilon}~.
  \label{kin-c}
\end{eqnarray}

One can not get longitudinal five brane by superposing orthogonal
membranes of the type (\ref{conf1-1}), which would have membrane
charges in it. But still there exist the solutions of longitudinal
five brane stretched along $X^{i'}$ directions without membrane
charges. It is realized by the configurations of the form
\begin{eqnarray}
X^{4}=x^4 \cos(\mu t/6) - x^7 \sin(\mu t/6) ~, \ \ \
X^{7}=x^4\sin(\mu t/6) + x^7 \cos(\mu t/6)~,\cr X^{5}=x^5 \cos(\mu
t/6) - x^8 \sin(\mu t/6) ~, \ \ \ X^{8}=x^5 \sin(\mu t/6) + x^8
\cos(\mu t/6)~~,
\label{conf2}
\end{eqnarray}
while all other coordinates vanish, i.e. $X^i, X^{i'}=0$. Here
$x^{i'}$ are time-independent and satisfy the condition
\begin{equation}
[x^{i'}, x^{j'} ] = \frac{1}{2}\epsilon_{i'j'k'l'}[x^{k'},
x^{l'}]~, \label{lc}
\end{equation}
where $\epsilon_{i'j'k'l'}$ is the Levi-Civita symbol in four
indices. The condition (\ref{lc}) gives rise to the condition
\begin{equation}
[X^{i'}, X^{j'} ] = \frac{1}{2}\epsilon_{i'j'k'l'}[X^{k'},
X^{l'}]~,
\end{equation}
in which the dynamical supersymmetry transformations reduce to
\begin{eqnarray}
\delta\theta &=& \frac{i}{2} [ X^{i'}, X^{j'}]
 \gamma_{i'j'}\epsilon~.
  \label{dyn-d}
\end{eqnarray}
Therefore it describes longitudinal five-brane filling four
directions in $X^{i'}$ and sticking at the origin in the
transverse directions. One can easily see from (\ref{dyn-d}) that
it preserves $\frac{1}{4}$ supersymmetry. It does not contain
membrane charges but has angular momentum on the $4-7$ and $5-8$
planes. Again this configuration is possible only in the large $N$
limit \cite{ban157}

Final configuration we will describe is the longitudinal five
brane filling $X^1, X^2, X^3$-directions and one direction in
$X^{i'}$. This five brane has infinite number of membranes
embedded ellipsoidally. One can take the coordinates of the form
\begin{equation}
X^{4}=r \cos(\mu t/6)~, \ \ \ X^{5}=r \sin(\mu t/6)~,
\end{equation}
and
\begin{equation}
X_{\pm} = X^1\pm i X^2 = e^{\pm i\mu t/6} x_{\pm}~,\ \ \ X^3 =
x^3~,
\end{equation}
where $r, x_{\pm}$, and $x^3$ are time independent. All other
coordinates should be taken to be zero. The configuration is
characterized by
\begin{equation}
[x^3, x_{\pm} ] = \pm \frac{\mu}{2 R}~, \ \ \ [x_+, x_-]=
\frac{2\mu}{3R} x^3~, \label{com}
\end{equation}
and
\begin{equation}
[x_\pm, r ] = i {\cal F}_{\pm}~, \ \ \
[x^3, r]= 0~,
\end{equation}
with ${\cal F}_- = {\cal F}_+^\dagger$. Note that the commutation
relations (\ref{com}) are not those of fuzzy sphere, rather they
are of fuzzy ellipsoid with
\begin{equation}
\frac{(X^1)^2}{6}+\frac{(X^2)^2}{6}+\frac{(X^3)^2}{4}=(\frac{\mu}{R})^2
j(j+1)
\end{equation}
for $(2j+1)$-dimensional representation of $SU(2)$. Therefore one
may regard this configuration as the one of longitudinal five
branes in which ellipsoidal membranes are embedded. For the
dynamical supercharges of the type
\begin{eqnarray}
\Pi \epsilon=  \gamma_{12}\epsilon =  \gamma_{45}\epsilon = \pm
i\epsilon~,
\end{eqnarray}
the dynamical supersymmetry transformation rule reduces to
\begin{eqnarray}
\delta\theta &=& \frac{i}{2} [ X_{\pm}, r]
 e^{\pm{i\mu \over 4} t}\gamma_{14}\epsilon~.
 \label{dyn-e}
\end{eqnarray}
Therefore the configuration has $\frac{1}{8}$ supersymmetry which
are generated by the combination of dynamical supersymmetry
transformation (\ref{dyn-e}) and kinematical supersymmetry
transformation (\ref{kin-c}).

\section{Discussions}
\label{conc}

One immediate question in exploring M/string theory on some
background is what kind of nonperturbative BPS objects survive in
that particular background. The same question has been explored in
the IIB superstring theory on the pp-wave background
\cite{bla242,met044,met109} in the works \cite{dab231,ske054} by
studying the boundary conditions of open strings on D-branes.\footnote{See 
also \cite{kns025,ak134} for the pp-wave background originating from 
$Ads_3\times S^3$ geometry.} On
the other hand, in the IIA superstring theory side, we rather have
to consider M theory on eleven dimensional pp-wave background.
Matrix theory, so far, is the only tractable candidate to describe
M theory. Hence, it seems natural to probe M theory on pp-wave
background using the corresponding massive matrix theory.

In this paper we analyzed supersymmetry algebra of the matrix
model and found some central charges which can be activated only
in the large $N$ limit. Then using the supersymmetry
transformation rules of fermions we identified some transverse
membrane solutions and longitudinal five brane solutions. They
exists only in the large $N$ limit and can not be static, but
rather they are rotating solutions with specific angular
frequency. The angular frequency is proportional to the mass $\mu$
and is thus due to the non-trivial pp-wave background. One may
wonder how it would be possible for the infinitely stretched
branes to have angular momentum. However one should note that, as
pointed out in \cite{das185}, the eleventh coordinate of the
pp-wave background geometry loose its meaning as a spatial
coordinate in the asymptotic region. This makes questionable using
the matrix theory as a proper description in that region. One may
hope quadratic potential terms confine $X$ in the finite region
giving effective cut-off. In any case our results give the
existence of such solution in general terms which would be
realized in the case when the transverse spaces are compactified.

It would be interesting to probe BPS branes in IIB superstring
theory \cite{dab231,ske054} in the context of IIB matrix string
theory \cite{gop174,bon213,ver059}.

After the completion of this paper, there appeared a paper on the
web \cite{sug070} which  is closely related to section \ref{alg}
in our paper.

\section*{Acknowledgments}
 The
work of S.H. was supported in part by grant No. 2000-1-11200-001-3
from the Basic Research Program of the Korea Science and
Engineering Foundation.

%------------------ References -------------------------

\end{document}